\documentclass[10pt,conference]{IEEEtran}

\usepackage{graphicx} 
\usepackage{float}

\usepackage{url}
\usepackage{subcaption}
\usepackage{dblfloatfix}
\usepackage{braket}
\usepackage{balance}
\usepackage{amsmath}

\title{A Unified Framework for Operator Backpropagation and Observable Measurement in Quantum Computing}

\date{}

\author{
\IEEEauthorblockN{Kevin Dougherty}
\IEEEauthorblockA{
Mathematics and Computer Science Division (MCS)\\
Argonne National Laboratory\\
Lemont, IL, USA\\
kdougherty@anl.gov
}
\and
\IEEEauthorblockN{Ji Liu}
\IEEEauthorblockA{
Mathematics and Computer Science Division (MCS)\\
Argonne National Laboratory\\
Lemont, IL, USA\\
ji.liu@anl.gov
}
}

\begin{document}

\maketitle

\begin{abstract}
Decoherence imposes severe limits on quantum circuit depth, motivating methods to reduce circuit depth. Operator backpropagation has recently been proposed as a way to lower circuit depth by classically backpropagating observables through subcircuits. While it lowers the circuit depth, it also significantly increases the number of backpropagated observables that must be measured. This calls for an efficient measurement protocol tailored to these backpropagated observables. Moreover, errors arise both during the backpropagation procedure and during measurement, and they accumulate in the final computation. No prior work has examined how to integrate operator backpropagation with advanced measurement protocols, and how to jointly optimize the combined process. 

In this paper, we introduce a comprehensive framework to determine the appropriate measurement protocol to achieve a desired level of accuracy with the fewest number of measurements based on the number of qubit-wise commuting (QWC) groups formed from operator backpropagation. We analyze and categorize the primary sources of error that arise through our workflow, including error incurred from the backpropagation algorithm, truncation error, and shot noise variation. We also examine truncation strategies to reduce the number of backpropagated observables by characterizing the structure of the set of backpropagated observables.
\end{abstract}


\section{Introduction}


Quantum computing has significant potential for solving problems in areas such as chemistry simulation~\cite{chemistry_VQE}, many-body physics~\cite{fauseweh2024quantum_manybody}, and combinatorial optimization~\cite{qaoa}. However, current devices are in the Noisy Intermediate-Scale Quantum (NISQ) era~\cite{preskill2018quantum_NISQ}, where system sizes are limited and operations are subject to noise without full error correction. This makes noise a major obstacle to the practical use of quantum algorithms. To mitigate these limitations, it is essential to reduce quantum circuit size and depth, even if additional classical computation is required. Beyond traditional quantum circuit optimization and synthesis, recent approaches aim to lower quantum cost by reformulating parts of the computation into hybrid quantum–classical workflows.
There have been many techniques developed to reduce the depth of quantum circuits (\cite{CPT}, \cite{LOWESA}, \cite{OBP}, \cite{PP_sim}) through classical simulation. Operator backpropagation~\cite{OBP} is one such method. It divides the original quantum circuit into two parts: one executed on a quantum device and the other simulated classically. The term backpropagation refers to propagating the observable backward through the subcircuit in the Heisenberg picture. While this approach reduces the quantum circuit depth, it increases the number of observables that must be measured, and the backpropagation process can generate significantly more backpropagated observables than in the original circuit.

Measuring expectation values of observables is critical in many quantum algorithms such as VQE~\cite{peruzzo2014variational_VQE} and in the operator backpropagation framework. A wide range of measurement protocols has been proposed to reduce the measurement overhead. Among these, two of the most commonly used approaches are measurement basis reduction via commuting groups~\cite{gokhale2019minimizing_commuting, zhao2020measurement_commuting, OBS_estimation} and classical shadow protocols~\cite{classicalshadows,derandomized,shallowshadow,triply_efficient}. 

It is critical to co-optimize the operator backpropagation process and the measurement protocol. During backpropagation, truncation error may arise when observables with small coefficients are discarded to avoid an exponential increase in the number of terms. The measurement stage also introduces shot noise due to the finite number of shots, and the scaling of this error depends on the chosen measurement protocol. Since errors from both stages accumulate and affect the final accuracy, a unified framework is needed to co-optimize these processes and evaluate how different backpropagation strategies interact with various measurement protocols.



\begin{figure*}[!t]
 \centering
        \includegraphics[width=0.9\textwidth]{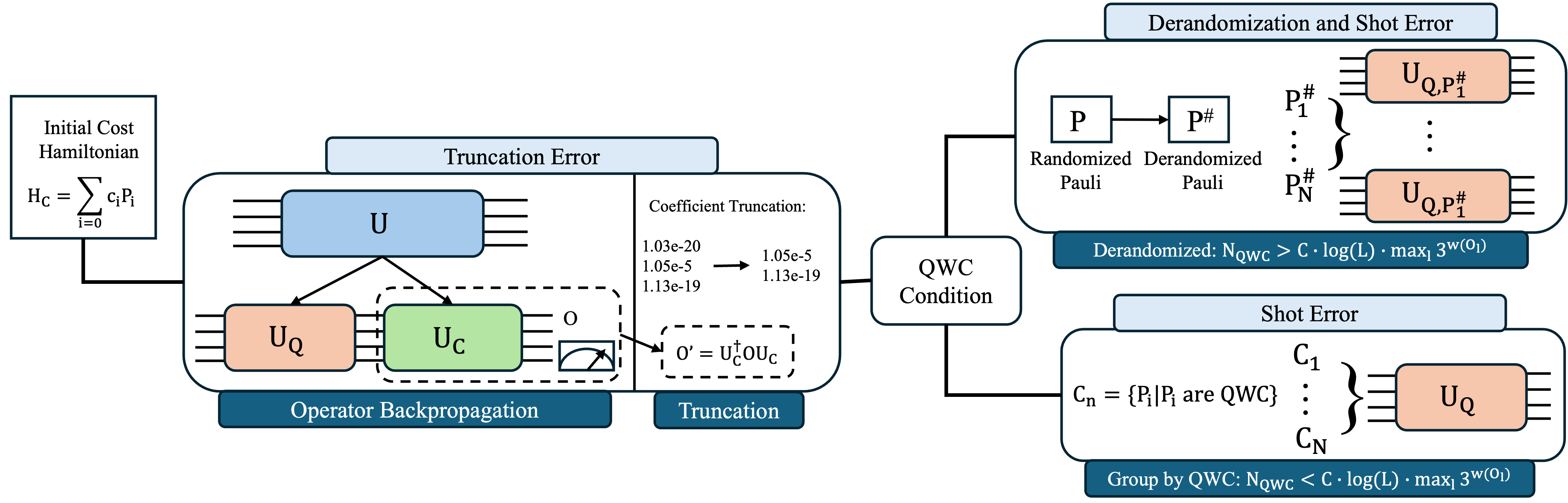}
    \caption{Comprehensive framework to determine the appropriate measurement protocol to achieve a desired level of accuracy with the fewest number of measurements based on the number of commuting groups generated from backpropagation.}
\label{fig:framework}
\end{figure*}

In this paper, we proposed a framework for operator backpropagation and observable measurement protocols. The framework accounts for the overall error rate and selects the measurement protocol according to the characteristics of the observables. Our study also examines the distribution properties of observables and provides useful insights into how they affect accuracy and cost. We evaluate the framework using a variety of circuits, including: Heisenberg Hamiltonian simulation, QAOA, Ising, and Quantum Fourier Transform (QFT)  circuits under different operator backpropagation depths. Our contributions are summarized as follows.

\begin{enumerate}
    \item Propose a framework for estimating the expectation values of observable with the fewest number of measurements based on the number of qubit-wise commuting (QWC) groups.
    \item Define an improvement ratio that quantifies the advantage of derandomization over direct measurement after backpropagation.
    \item Perform a comprehensive error analysis on the primary aspects of our workflow.
    \item Study the truncation algorithms for propagating a quantum circuit to characterize the structure of the backpropagated observables.
\end{enumerate}


\section{Background}
\label{sec:background}
This section will provide an overview of existing propagation methods, along with certain truncation strategies, and some of the observable measurement protocol literature. It will also establish why we chose to specifically combine operator backpropagation, with the derandomized algorithm over the other propagation and measurement methods presented.

\subsection{Propagation Methods}
\label{sec:propagation methods}
The goal of propagating an observable through a quantum circuit is to offload portions of the quantum circuit to a classical simulator. This reduction in the depth of the quantum circuit has the potential to reduce the error incurred by running the quantum circuit. Various implementations of Pauli propagation have been proposed, such as operator backpropagation (OBP) \cite{OBP}, and the low-weight efficient simulation algorithm (LOWESA) \cite{LOWESA}.

While these implementations may differ in other regards, the general framework to propagate an observable through a circuit, is as follows. Start with an initial observable $O$, and evolve it through the gates of the circuit starting at the end of the circuit. Consider a quantum circuit $U$, which we divide into two subcircuits $U_c$ and $U_q$ such that $U = U_cU_q$. Our goal is to estimate the expectation value $\braket{O}$ of an observable $O$ with respect to the quantum state $\ket{\psi}$ prepared by the circuit $U$. This leads to:
\begin{equation}
\braket{O} = \bra{0}U^{\dagger}OU\ket{0} = \bra{0}U_q^{\dagger}U_c^{\dagger}OU_cU_q\ket{0} = \bra{0}U_q^{\dagger}O'U_q\ket{0}
\end{equation}

Thus, the task reduces to running a subcircuit $U_q$ on the quantum device and measuring the new observable $O' = U^\dagger_cOU_c$. This new observable is obtained through classical calculation and expands into a linear combination of Pauli observables:
\begin{equation}
    O' = U^\dagger_cOU_c  = \sum \alpha_iP_i
\end{equation}
where $\alpha_i$ are the coefficients of the backpropagated observables and $P_i \in \{I, X, Y, Z\}^{\otimes n}$ are multi-qubit Pauli operators, which we call our set of backpropagated observables~\cite{OBP}. The number of backpropagated observables grows exponentially with the number of non-Clifford operations in $U_c$. To address this exponential overhead, truncation strategies have been proposed and will be discussed in the following section.


More specifically, OBP~\cite{OBP} relies on Clifford Perturbation Theory~\cite{CPT}, thus works particularly well for circuits with many Clifford or near-Clifford gates. LOWESA creates a classical surrogate of the circuit landscape, and while the initial overhead cost to run the algorithm is high, it can be subsequently run efficiently on the target observables~\cite{LOWESA}. One of the primary bottlenecks of LOWESA is the number of backpropagated observables that are generated, which is significantly higher than that of OBP, and will quickly cause a computer to run out of memory. OBP appears to generate more accurate results with a fewer number of observables. Another related work is Pauli Propagation~\cite{pp}, which follows a similar propagation process but is designed for simulating full quantum circuits rather than reducing circuit depth. As a result, our experiments were performed using the OBP framework provided on Qiskit~\cite{QiskitAddonOBP}.

\begin{figure*}[!t]
    \centering
    
    \begin{subfigure}[b]{0.3\textwidth}
        \centering
        \includegraphics[width=\textwidth]{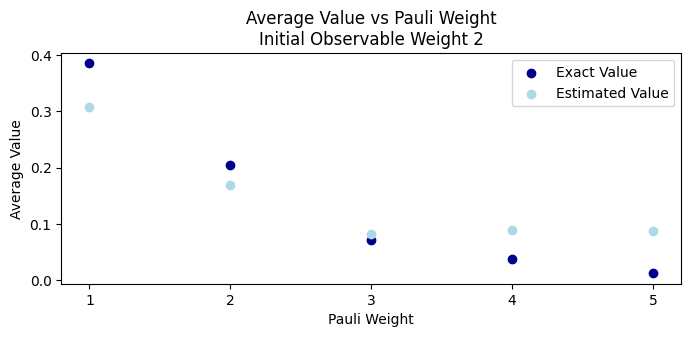}
        \caption{}
    \end{subfigure}
    \hfill
    \begin{subfigure}[b]{0.3\textwidth}
        \centering
        \includegraphics[width=\textwidth]{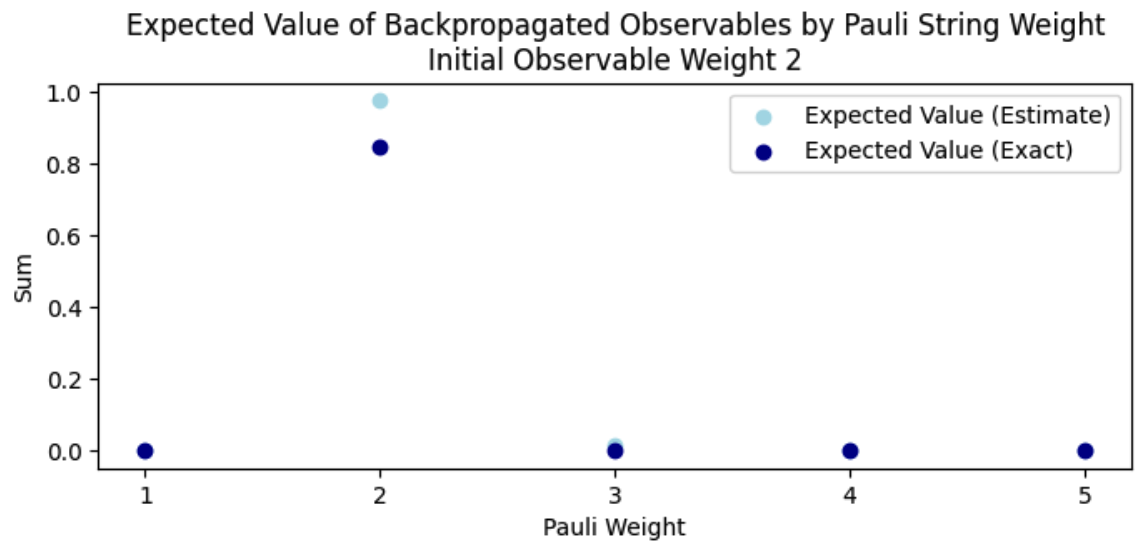}
        \caption{}
    \end{subfigure}
    \hfill
    \begin{subfigure}[b]{0.3\textwidth}
        \centering
        \includegraphics[width=\textwidth]{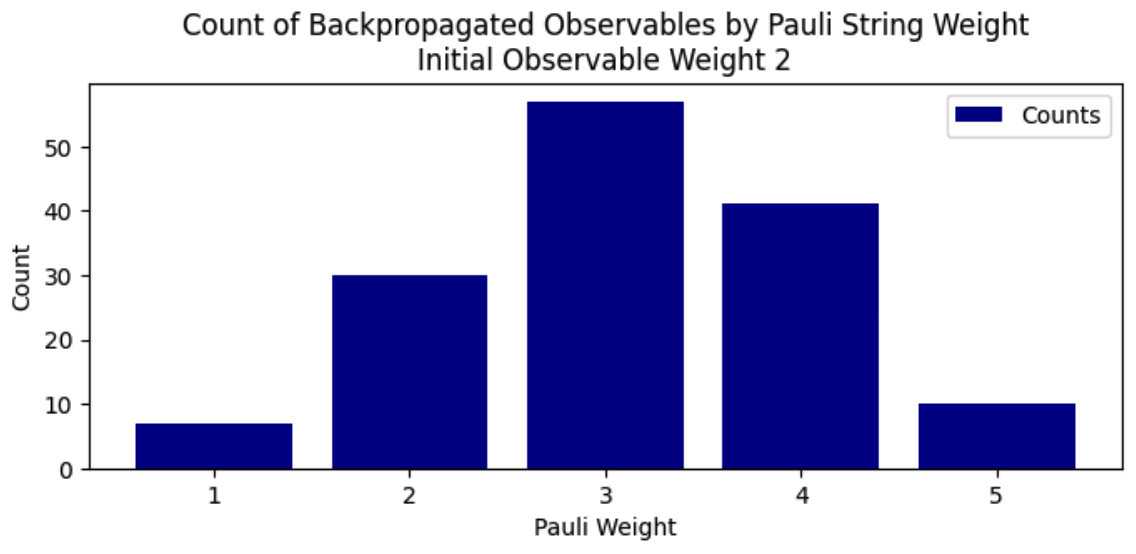}
        \caption{}
    \end{subfigure}

    \vspace{0.2cm}

    \begin{subfigure}[b]{0.3\textwidth}
        \centering
        \includegraphics[width=\textwidth]{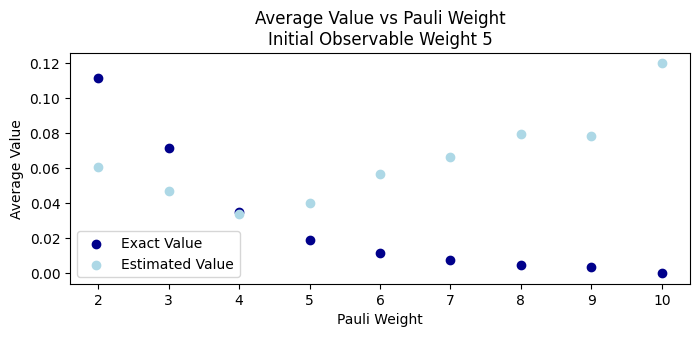}
        \caption{}
    \end{subfigure}
    \hfill
    \begin{subfigure}[b]{0.3\textwidth}
        \centering
        \includegraphics[width=\textwidth]{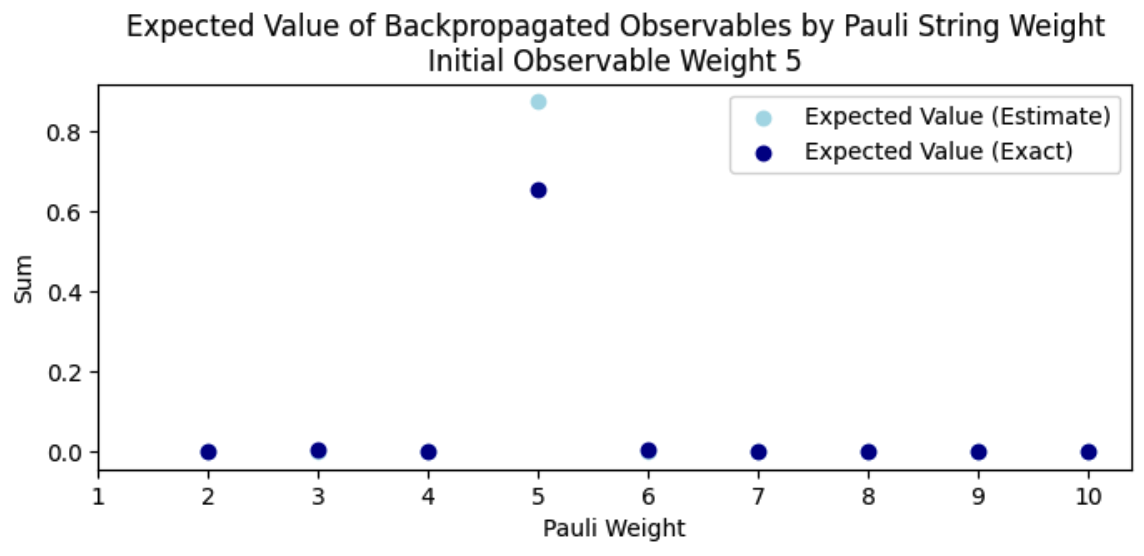}
        \caption{}
    \end{subfigure}
    \hfill
    \begin{subfigure}[b]{0.3\textwidth}
        \centering
        \includegraphics[width=\textwidth]{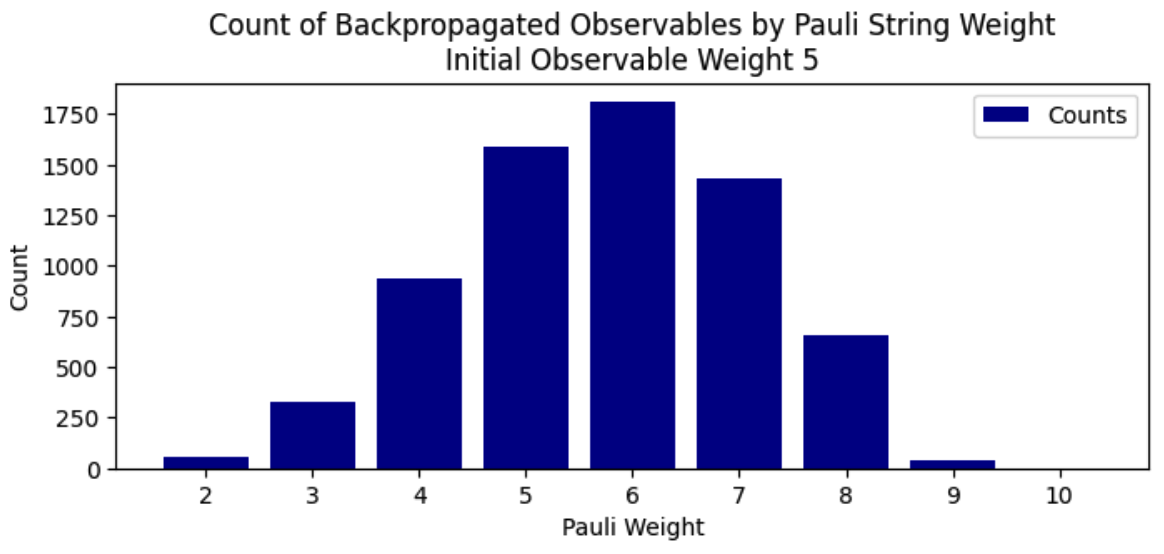}
        \caption{}
    \end{subfigure}

    \vspace{-0.2cm}
    \caption{Analyzing the distribution of the Pauli weights of the backpropagated observables using a XYZ Heisenberg Hamiltonian}
    \vspace{-0.2cm}
    \label{fig:pauli_weight}
\end{figure*}

Operator backpropagation is typically implemented with an error budget, and allows for truncating Pauli strings with small coefficients. Several truncation strategies have been proposed. The OBP framework adopts coefficient-based truncation, removing Pauli terms with small coefficients. The Pauli Propagation package employs a weight-based strategy that truncates high-weight Pauli strings. This approach is based on the assumption that, for the ground state $\ket{0}$ the probability of a Pauli observable having a nonzero expectation value decreases exponentially as the Pauli weight increases. This assumption holds when simulating the full circuit but may not hold when measuring after a shorter subcircuit, where the state is $U_q\ket{0}$.  Therefore, experimental evaluation is needed to assess these truncation strategies.

\subsection{Observable Measurement Protocols}

Various measurement protocols have been proposed to reduce the overhead of measuring expectation values of observables. The standard approach is to perform measurement basis reduction using commuting groups~\cite{zhao2020measurement_commuting, gokhale2019minimizing_commuting, OBS_estimation}. Since commuting Pauli observables can be measured in the same basis, the problem becomes finding the maximum commuting clique in the set of Pauli terms. Different strategies have been proposed, ranging from qubit-wise commuting (QWC) groups to general commuting groups~\cite{gokhale2019minimizing_commuting}. The OBP framework leverages QWC groups for reducing the number of measurement basis.

Shadow tomography is another class of measurement protocols based on randomized measurements. These measurements allow for the efficient estimation of the expected value $Tr(O_i\rho)$ for many observables $O_i$~\cite{shadowtomography}. The original classical shadows method forms a sketch of the classical shadow $S_p$ by applying random Clifford unitaries $U$ to the state $\rho$ to measure $U \rho U^\dagger$ in the computational basis~\cite{classicalshadows}. There have been several updated versions of the classical shadows procedure since its inception. One such version, robust shallow shadows, is a two-step procedure where step one involves learning the noise of a quantum circuit and step two involves preparing the application state $\rho$, finding a matrix product state representation for the observables of interest, and inferring the values of the observables~\cite{shallowshadow}. Another variation of classical shadows is a derandomization algorithm, which replaces random single qubit measurements with fixed Pauli measurements~\cite{derandomized}. This can significantly reduce the sample complexity, or the number of measurements needed, and improve the accuracy of the results over the initial classical shadows procedure, particularly for high-weight Pauli observables~\cite{derandomized}. The commuting-group strategy and shadow-based strategies each offer advantages and limitations. In our framework, we incorporate both QWC and the derandomized protocol from these two families of measurement strategies. The choice of protocol is determined based on the properties of the observables.

\subsection{Compatibility Between Methods}

Combining these propagation techniques with the shadow procedures is natural because the primary bottleneck incurred from propagating an observable is the potential exponential growth in the observable size, and the shadow procedures are able to compensate for this growth by only needing a logarithmic number of measurements. As theoretically demonstrated in \cite{qsim}, the combination of shadow tomography with Pauli propagation only requires $polylog$ number of measurements.

The derandomization procedure adds a nominal amount of depth to the backpropagated circuit, and is capable of more accurately measuring some of the higher-weight Pauli strings that may be incurred with coefficient-based truncation. 
Other shadow tomography methods, like \cite{shallowshadow}, may add a significant amount of depth to the circuit, thus would not be an ideal choice if the goal is to shorten the depth of the circuit with backpropagation.

\begin{figure*}[!t]
    \centering
    \includegraphics[width=0.88\textwidth]{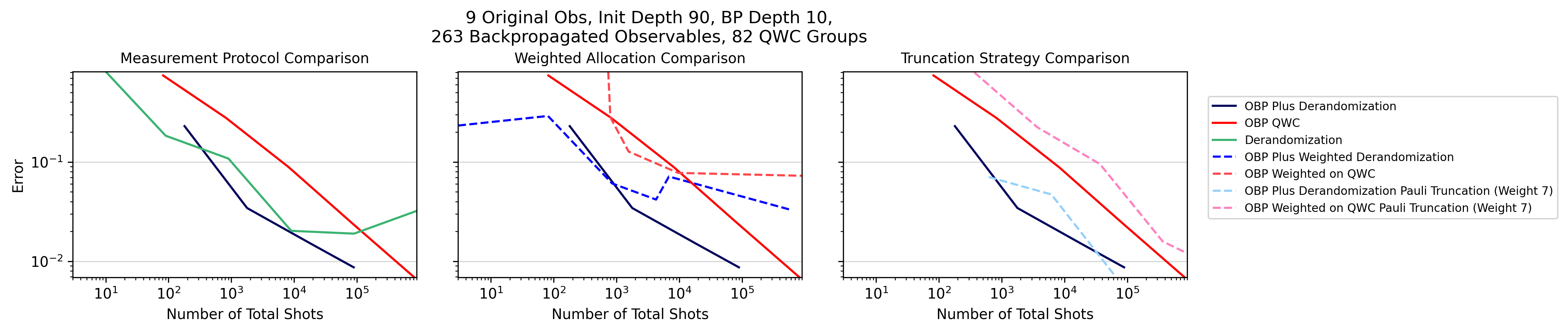}
    \caption{\textbf{Operator Backpropagation-Derandomization Protocol Results on a XYZ Heisenberg Hamiltonian Circuit.}}
    \label{fig:XYZresults}
\end{figure*}

\begin{figure*}[!t]
    \centering
    \includegraphics[width=0.88\textwidth]{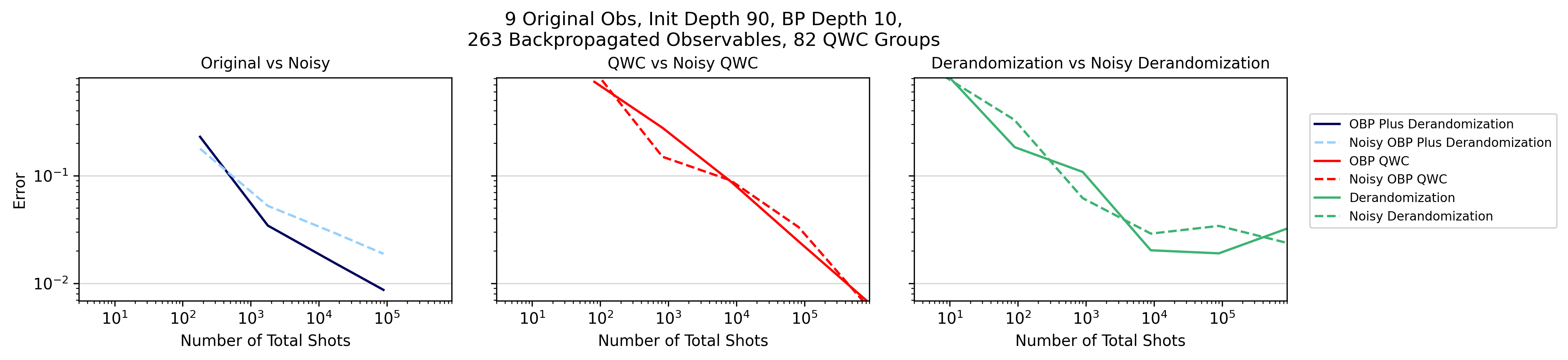}
    \caption{\textbf{Operator Backpropagation-Derandomization Protocol Noisy Results on a XYZ Heisenberg Hamiltonian Circuit.}}
    \label{fig:noisyXYZresults}
\end{figure*}

\section{Methodology}
\label{sec: methodology}

The framework is defined by two coupled design choices: how the total error budget is allocated between OBP and the measurement protocol, and how the measurement protocol is selected given this allocation. The measurement protocol is selected by minimizing the total number of shots required to achieve a target accuracy. This shot cost depends on three factors: the number of QWC groups, the number of backpropagated observables, and the maximum weight of any individual backpropagated observable, which is controlled with the truncation strategies. The total error budget allocated to OBP and the measurement protocol is determined by the choice of truncation strategy and by the number of shots used to suppress readout noise.

\subsection{Truncation}
\label{sec: truncation}
We consider two truncation strategies, which may be applied independently or in combination. The strategies directly impact the error introduced in OBP and the derandomization procedure in the measurement protocol. Specifically, backpropagated observables may be truncated by coefficient magnitude or Pauli weight.

\subsubsection{Truncation strategy case study}

In this Section, we examine the Pauli-weight distribution of representative problems in Hamiltonian simulation. The results indicate that high-weight Pauli strings typically carry small coefficients, supporting the use of Pauli-weight–based truncation strategies. 

Figure~\ref{fig:pauli_weight} analyzes the distribution of the weights of the backpropagated observables. Data on the left side of the Figure in Panels (a, c, d) was run with \textit{ZZIIIIIIII} and data on the right side of the figure in Panels (b, d, f) was run with \textit{ZZZZZIIIII}. Panels (a) and (b) show the estimated and exact average absolute expected value of the backpropagated observables for varying Pauli weights. This numerically demonstrates that higher weight initial observables will generate higher weight backpropagated observables, and that backpropagated observables with high Pauli weights tend to have an expected value near 0, especially compared to low-weight Paulis, thus demonstrating the validity of truncating backpropagated observables by Pauli weight. Panels (c) and (d) show the sum of the expected values of the backpropagated observables with a given Pauli weight. In both figures, the expected value appears to be almost exclusively determined by the Paulis with the weight of the initial observable. This is due to the size of the coefficient of the initial observable in the backpropagated set of observables. During backpropagation, the Pauli with the largest coefficient, by some order of magnitudes, is identical to the initial observable. Panels (e) and (f) show the distribution of the backpropagated observables by Pauli weight.

\subsection{QWC Condition}
\label{subsec: QWC Condition}

A crucial insight of our framework relies on Theorem 1 in \cite{derandomized}, which states that the number of measurements $M$ required for predicting $L$ Pauli expectation values up to additive error $\epsilon$ using their randomized protocol is proportional to the $log$ of the number of observables times 3 to the maximum weight of the observables:

\begin{equation}
M \propto \frac{\log(L) \cdot \max_\ell 3^{w(O_\ell)}}{\varepsilon^2}.
\end{equation}

The derandomized algorithm is guaranteed to perform at least as favorably as the average case of the randomized protocol, but the actual difference can be much more pronounced. For a conservative comparison with QWC grouping, we use the randomized bound as an upper limit on the measurement budget. 

In contrast, directly measuring a set of observables grouped into $N_{QWC}$ groups requires measurements proportional to the number of groups times the standard shot error, which is defined by the Bernoulli distribution. Assuming a worst case scenario where $p=0.5$, the total number of measurements is:

\begin{equation}
M_{QWC} \propto \frac{N_{QWC}}{4\varepsilon^2}
\end{equation}

Comparing these two approaches, the derandomized algorithm requires fewer measurements when:

\begin{equation}
N_{QWC} > C \cdot \log(L) \cdot \max_\ell 3^{w(O_\ell)},
\end{equation}

where $C$ is the proportionality constant from Theorem 1. A precise value of $C$ depends on the distribution of $\max_\ell 3^{w(O_\ell)}$ across the backpropagated observables, which in turn depends on the circuit and truncation strategy. Using the average of this distribution for each of our tested circuits respectively, our empirical tests find $C \approx 2.02$; however, a small sensitivity analysis in Table~\ref{tab:c-sensitivity} shows this value varies across circuit families. As a general principle, we find that when $L \gg N_{QWC}$, OBP with QWC is more efficient. Figure ~\ref{fig:pauli_weight} illustrates this distribution for the Heisenberg Hamiltonian, showing that high-weight observables contribute negligibly to the expectation value. When this condition holds, the derandomized algorithm should be applied. Otherwise if this condition fails, direct measurement by grouping the QWC groups is more efficient.

\begin{table}[H]
\caption{Per-family empirical averages and the crossover constant $C_{\mathrm{eff}} = N_{QWC}/(\log L \cdot 3^{w})$ using the average Pauli weight, weighted by coefficient magnitude.}
\label{tab:c-sensitivity}
\centering
\begin{tabular}{l|c|c|c|c}
\hline
\textbf{Metric} & \textbf{Heis 10q} & \textbf{Heis 40q} & \textbf{Ising 26q} & \textbf{QFT 18q} \\
\hline
Num Obs        & 36      & 3     & 1    & 6      \\
$L$            & 1674.08 & 80.33 & 4.00 & 370.67 \\
$N_{QWC}$      & 276.11  & 35.00 & 2.00 & 8.33   \\
Avg Weight     & 2.39    & 2.72  & 3.22 & 2.44   \\
\hline
$\mathbf{C_{\mathrm{eff}}}$ & \textbf{2.54} & \textbf{0.36} & \textbf{0.04} & \textbf{0.11} \\
$C_{\mathrm{std}}$ & 0.89 & 0.10  & --   & 0.07   \\
\hline
\end{tabular}
\end{table}

Not only is the decision to measure directly or to use the derandomization directly based on Equation 4, an improvement ratio, which yields the improvement of using the derandomization algorithm after backpropagating can be defined as:

\begin{equation}
\eta = \frac{N_{\text{QWC}}}{C \cdot \log(L) \cdot \max_\ell 3^{w(O_\ell)}}
\end{equation}

The reciprocal of this ratio is the improvement in directly measuring. In general, increasing the number of initial observables and decreasing the depth of the backpropagated circuit both result in a larger set of backpropagated observables. Depending on the structure of the initial observables, a larger set of backpropagated observables implies more QWC groups. The number of shots needed for OBP is the number of QWC groups formed by the backpropagated observables. The number of shots needed for the combined OBP-shadow protocol is logarithmic with respect to the number of backpropagated observables. The difference in the number of shots will therefore be the difference in these numbers as they scale.

\subsubsection{Optimal Shot Allocation}
Based on the truncation analysis, a natural inclination is to weight the number of shots based on the size of the coefficients. This is exactly what the OBP Weighted on QWC benchmark demonstrates in Figure ~\ref{fig:XYZresults}. The number of shots to use is determined by the size of the coefficients in their respective QWC groups. We use the shot allocation found in \cite{OBS_estimation}, which provides an initial shot allocation for each circuit based on the total available shot budget:

\begin{equation}
s_i = \operatorname{round}\!\left( 
S_{\text{total}} \frac{v_i}{\sum_{j=1}^N v_j}
\right),
\end{equation}

The largest allocated shot count is then increased to match the total shot allocation:

\begin{equation}
s_{\text{max}} = 
s_{\text{max}} + (S_{\text{total}} - \sum_is_i).
\end{equation}

However, one caveat which connects back to the discussion in Section~\ref{sec: truncation} on truncation methods, is that the distribution of weights is skewed towards the QWC group with the initial observable, and many of the QWC groups with small coefficients are allocated 0 shots. To rectify this, we allocate a minimum of 1 shot per group, which results in smaller shot allocation budgets being exceeded.

\subsection{Experiments}
\label{sec:experiments}

We performed our tests on a variety of circuits relevant to problems on near-term quantum devices because we were interested in exploring how this framework could be used to enhance the measurement protocol for distinct classes of circuits.

To perform our experiments for the VQE class of problems, we considered an XYZ Heisenberg Hamiltonian of the form:
\begin{equation}
\begin{split}
\hat{H} ={} &
\sum_{(j,k)\in E}
\bigl( J_x \sigma_j^x \sigma_k^x 
     + J_y \sigma_j^y \sigma_k^y 
     + J_z \sigma_j^z \sigma_k^z \bigr) \\
&+ \sum_{j\in V}
\bigl( h_x \sigma_j^x 
     + h_y \sigma_j^y 
     + h_z \sigma_j^z \bigr).
\end{split}
\end{equation}
The corresponding circuit is the same circuit used in \cite{OBP}.

\begin{figure*}[!t]
    \centering
    
    \begin{subfigure}[b]{0.24\textwidth}
        \centering
        \includegraphics[width=\textwidth]{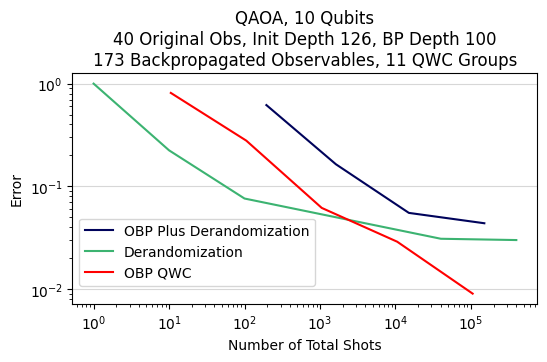}
        \caption{}
    \end{subfigure}
    \hfill
    \begin{subfigure}[b]{0.24\textwidth}
        \centering
        \includegraphics[width=\textwidth]{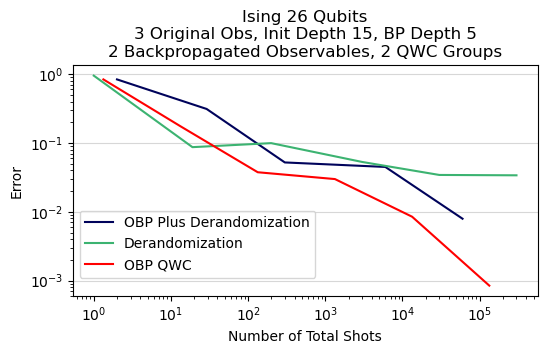}
        \caption{}
    \end{subfigure}
    \hfill
    \begin{subfigure}[b]{0.24\textwidth}
        \centering
        \includegraphics[width=\textwidth]{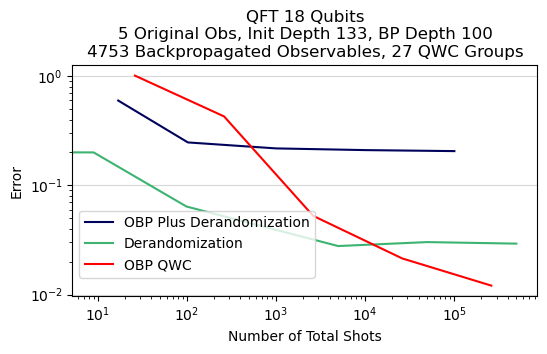}
        \caption{}
    \end{subfigure}
    \hfill
    \begin{subfigure}[b]{0.24\textwidth}
    \centering
    \includegraphics[width=\textwidth]{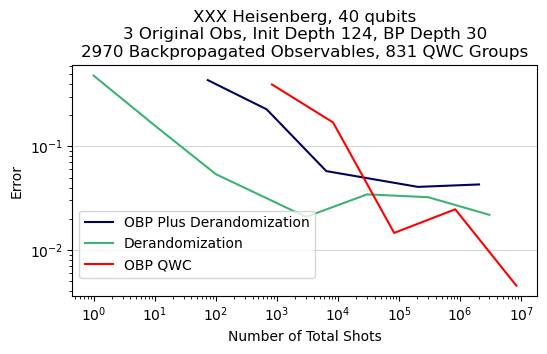}
    \caption{}
    \end{subfigure}
    \hfill
    
    \caption{\textbf{Evaluation Results of Operator Backpropagation-Derandomization Protocol on: QAOA, Ising, and QFT Circuits.}}
    \label{fig:QAOA_results}
\end{figure*}


For the QAOA class of problems, we considered a k-connected graph, using 10-vertex regular graphs with degree 6 and 8. To ensure that we were device agnostic, we randomly selected the circuit parameters, and used a universal basis set for QAOA.


The medium-sized circuit configurations from the QASMBench benchmark suite were used for both the Ising model and QFT circuits.

We use a modified version of Qiskit's Operator Backpropagation Addon. The standard implementation terminates backpropagation after a fixed number of QWC groups; we instead terminate at a specified circuit depth. This allows multiple observables to be backpropagated through the same circuit, yielding a single measurement scheme that the derandomized algorithm can apply across the full observable set. The trade-off is a larger backpropagated observable set, which is acceptable given the shared measurement budget.

For the purposes of experimenting, we focused on two-qubit Pauli observables. The tests on the XYZ Heisenberg Trotterized circuit used nine repetitions, resulting in a depth of 90. We set a max error per slice of 0.0001 for the backpropagation's truncation budget as this results in a larger set of backpropagated observables to demonstrate our results. We used Qiskit's Lima's Noise Model for our noisy measurement results.

\section{Evaluation}
\label{sec:evaluation}
We compare the combined framework against OBP and derandomization separately, examining (i) scaling with the number of initial observables, (ii) weighted shot allocation, (iii) the choice of truncation strategy, (iv) noisy results, (v) and generalization across circuit families.

\begin{table}[H]
\caption{Realized total measurements at the crossover, where $\epsilon$ is the
best error derandomization achieves for each family. Ratio is $M_{QWC}/M_{derand}$:
$>1$ means derandomization is cheaper, $<1$ means QWC is cheaper.}
\label{tab:crossover}
\centering
\begin{tabular}{l|c|c|c|c}
\hline
\textbf{Metric} & \textbf{Heis 10q} & \textbf{Heis 40q} & \textbf{Ising 26q} & \textbf{QFT 18q} \\
\hline
$\epsilon$         & 0.0077  & 0.0406 & 0.0079 & 0.2055 \\
$M_{\text{derand}}$ & 97948   & 204000 & 59999  & 100799 \\
$M_{\text{QWC}}$    & 2819468 & 70622  & 21959  & 1657   \\
\hline
\textbf{Ratio}              & \textbf{28.79$\times$} & \textbf{0.35$\times$} & \textbf{0.37$\times$} & \textbf{0.02$\times$} \\
\hline
\end{tabular}
\end{table}

\textbf{Scaling with initial observables.} We tested how the combined protocol scales with the number of initial observables for the 10-qubit Heisenberg Hamiltonian. We use approximately $10^5$ total shots, where shot noise and statistical variation are minimized. The combined protocol achieves a 1.93× improvement over OBP at 3 initial observables, rising to 8.37×, 11.28×, and 28.79× at 9, 18, and 36 observables (~\ref{tab:crossover}). This is approximately linear scaling with the number of initial observables.

\textbf{Weighted Allocation Comparison.} Figure ~\ref{fig:XYZresults}, Panel (b) shows a comparison using weighted allocations. Because the initial observable in the set of backpropagated observables dominates the expectation value, the weighted allocations perform worse relative to the standard protocols.

\textbf{Choice of truncation strategy.} Figure~\ref{fig:XYZresults}, Panel (c) demonstrates no noticeable difference between coefficient-based truncation and Pauli-weight truncation for the combined protocol. This is consistent with the observation that high Pauli-weight terms contribute negligibly to the total expectation value after backpropagation, as shown in Figure ~\ref{fig:pauli_weight}. Because the trends when truncating by Pauli weight rather than coefficient magnitude are qualitatively similar, the truncation by Pauli weight is omitted from Figure~\ref{fig:QAOA_results} for clarity.

\textbf{Noisy results.} Figure ~\ref{fig:noisyXYZresults} illustrates that the noisy results generally scale well across measurement protocols. The noisy results for the combined protocol, appear to trail off slightly, potentially converging towards the shadow plateau.

\textbf{Generalization across circuit families.} Figure~\ref{fig:QAOA_results} demonstrates the framework's applicability across QAOA, Ising, and QFT circuits. Within a circuit family, the criterion selects a consistent branch. For QAOA, Ising, and QFT, C\_eff lies well below the threshold across observables (Table~\ref{tab:c-sensitivity}), so QWC is selected. The 10 qubit Heisenberg sits around the crossover where derandomization performs better. Based on these results, it appears when there are many observables on a shallow subcircuit, derandomization wins, while QWC measurement dominates at scale (~\ref{tab:crossover}).

\textbf{Scaling to larger system sizes.} Figure~\ref{fig:QAOA_results} also reports results on a 40-qubit Heisenberg circuit with 2970 backpropagated observables and 831 QWC groups. In this case, OBP with QWC outperforms the combined protocol, demonstrating that the framework extends to larger system sizes, as the QWC condition holds for the 40 qubit Heisenberg Hamiltonian.


\section{Conclusion/Outlook}
\label{sec:conclusion}
In this paper, we present a framework for systematically determining whether to measure backpropagated observables directly or through derandomization, a decision that significantly impacts measurement efficiency in VQAs. Through comprehensive error analysis, we identify the key factors governing this trade-off: the number of QWC groups, the logarithmic scaling of derandomization with observable count, and the exponential dependence on Pauli weight. We define an improvement ratio that enables practitioners to make informed decisions about measurement strategies for their specific cases. Future work may involve experimenting with increasing the depths of backpropagated circuits to improve commutativity for direct measurements, or increasing the depths of the backpropagated circuits to improve a shadow method.

\section*{Acknowledgment}
This material is based upon work supported by the U.S. Department of Energy, Office of Science, National Quantum Information Science Research Centers. This material is also based upon work supported by the DOE-SC Office of Advanced Scientific Computing Research MACH-Q project under contract number DE-AC02-06CH11357.

\clearpage
\balance
\bibliographystyle{IEEEtran}
\bibliography{references}

\vfill

\small

\framebox{\parbox{\linewidth}{
The submitted manuscript has been created by UChicago Argonne, LLC, Operator of 
Argonne National Laboratory (``Argonne''). Argonne, a U.S.\ Department of 
Energy Office of Science laboratory, is operated under Contract No.\ 
DE-AC02-06CH11357. 
The U.S.\ Government retains for itself, and others acting on its behalf, a 
paid-up nonexclusive, irrevocable worldwide license in said article to 
reproduce, prepare derivative works, distribute copies to the public, and 
perform publicly and display publicly, by or on behalf of the Government.  The 
Department of Energy will provide public access to these results of federally 
sponsored research in accordance with the DOE Public Access Plan. 
http://energy.gov/downloads/doe-public-access-plan.}}

\end{document}